\documentclass[11pt]{amsart} 
\usepackage[english]{babel}
\usepackage{amsmath,amssymb,amscd,amsthm} 
\usepackage{subcaption}
\usepackage{graphicx}
\usepackage[dvipsnames]{xcolor}
\usepackage{listings}
\usepackage[utf8]{inputenc}
\usepackage{hyperref}
\usepackage{array, multirow}
\usepackage{lipsum}
\usepackage[sort, numbers]{natbib}

\usepackage[inner=2 cm, outer = 2 cm, top=2 cm, bottom = 2 cm]{geometry}
\usepackage{setspace}

\hypersetup{
	colorlinks=true,
	linkcolor=blue,
	filecolor=magenta,      
	urlcolor=blue,
	citecolor=black
}

\begin{document}
\setstretch{1.25}

\title{Stochastic epidemic model on a simplicial complex}

\author[G. Palafox-Castillo \& A. Berrones-Santos]{Gerardo Palafox-Castillo and Arturo Berrones-Santos}

\maketitle 
\begin{abstract}
Complex networks with pairwise connections have been vastly used for the modeling of interactions within systems. Although these type of models are capable of capturing rich structures and different phases within a great variety of situations, their lack of explicit higher order interactions might result, in some contexts, limited. In this work a stochastic epidemic model on a simplicial complex is defined, generalizing the known Markovian SIR epidemic process on networks. 
The stochastic microscopic process is studied 
by direct simulations and a homogeneous mean field description is 
developed. The simple dissipative SIR infection dynamics permits
a thorough characterization of the epidemic for arbitrarily high order interactions.
\end{abstract}
 

\section{Introduction}\label{sect:intro}
The mathematical abstraction of large interacting systems by networks has a long history, yet they have the natural limitation of only capturing pairwise interactions directly. Higher-order network models allow us to treat a finite group of ``individuals'' (nodes) as a single entity (in our case, a simplex). Among the many uses of traditional network systems is the modelling of epidemics through compartmental models \citep{Newman_2002, Pastor-Satorras_Vespignani_2015, vanmieghem2014exact, B2020}, where nodes are divided in compartments (e.g. susceptible, infectious) and spread a virus, rumour, disease, etc., to its neighbours. Common compartmental models include Susceptible - Infectious (SI), Susceptible - Infectious - Susceptible (SIS), Susceptible - Infectious - Recovered (SIR), Susceptible - Infectious - Recovered - Susceptible (SIRS). In the present work, a well known \citep{B2020} Markovian SIR model for networks is generalized to a simplicial complex. Epidemic processes on simplicial complexes have been considered before, in their SIS \citep{Cisneros-Velarde_Bullo_2020, Iacopini_Petri_Barrat_Latora_2019, Matamalas_Gomez_Arenas_2020} and SIRS variants \citep{wang2021simplicial}. Some reasons behind using simplicial complexes and not hypergraphs is to take advantage of the tools applied algebraic topology and topological data analysis may offer to better understand the subjacent structure. As mathematical objects, simplicial complexes are well studied \citep{Fomenko_Fuchs_2016, Munkres_1984}. Simplicial complexes have also been used to model non-Markovian higher-order network models where indirect influence between nodes through transitive paths is not assumed \citep{Lambiotte_Rosvall_Scholtes_2019}. There have been other applications of simplicial complexes to network science \cite{AAF2019, Kartun-Giles_Bianconi_2019, Zhou_Maletic_Zhao_2018}, complex systems in general \citep{Salnikov_Cassese_Lambiotte_2019}, and opinion dynamics such as consensus models \citep{deville2021consensus} or voter models \citep{horstmeyer2020adaptive}. Other stochastic processes, such as percolation, have also been considered on simplicial complexes \citep{zhao2022higher, bianconi2019percol}.

\subsection{Preliminaries}
An abstract simplicial complex $X$ with set of nodes $V$ is a collection of subsets of $V$ such that:
\begin{enumerate}
    \item for all $v \in V$: $\{ v \} \in X$;
    \item if $\sigma \in X$ and $\sigma^* \subseteq \sigma$ then $\sigma^* \in X$.
\end{enumerate}
Elements of $X$ are called simplices. A simplex containing $k+1$ nodes is called a $k$-simplex, and it is said to have dimension $k$. The dimension of a simplicial complex $X$ is the maximum dimension of the simplices it contains. 
A clique of a graph $G$ is a complete subgraph of $G$. Cliques with $k$ nodes will be called $k$-cliques. The clique complex of a graph $G$ is the simplicial complex $X = \{ \sigma : \sigma \hbox{ is a clique in } G\}$. The $d$-dimensional clique complex of a graph is the simplicial complex 
\begin{equation*}
    X = \bigcup_{k = 1}^{d+1} \{ \sigma: \sigma \hbox{ is a $k$-clique in } G\}.
\end{equation*}
The goal is to showcase the impact of higher order interactions in a contagion process by generalizing a stochastic SIR epidemic model from networks to simplicial complexes. In particular, the following continuous time Markovian epidemic on a network is generalized: at time $t= 0$ all nodes in the network are susceptible, except a randomly chosen one which is infectious. While a node is infectious, it has infectious contacts with each susceptible neighbor in the network randomly in time according to independent Poisson processes with rate $\beta$. Each infected individual remains infectious for a period $I$ exponentially distributed with mean $1 / \gamma$, after which it recovers and becomes immune. All infectious periods and contact processes are defined independently. The epidemic goes on until the first time $T$ that there are no infectious individuals and the epidemic stops. This definition and more work on epidemic processes on social networks can be found in the work by \citet{B2020}.

To construct the random simplicial complexes mentioned in Section \ref{sect:mean_field}, the model given by \citet{Iacopini_Petri_Barrat_Latora_2019} was used, with the exception that here we conditioned on the connectedness of the subyacent Erd\H{o}s-Rényi graph instead of taking the largest component. To be explicit, the random simplicial complexes in Section \ref{sect:mean_field} were built by taking an Erd\H{o}s-Rényi graph $G(n, p_1)$ and then, only if this was connected, adding every three-vertex clique with probability $p_2$ as a 2-simplex, with the edges as 1-simplices. This is similar to the model by \citet{Costa2016}, with the difference that the latter conditions on the existence of the necessary edges when adding a 2-simplex to the simplicial complex.

The outline of the work is the following. In Section \ref{sect:simplicial_sir} the stochastic SIR process on an arbitrary simplicial complex of dimension $d$ is defined. In order to compare this process to the traditional stochastic SIR on networks, in Section \ref{sect:experiments} a  stochastic, SIR epidemic is considered on a network and on its associated 2-clique complex. The goal is to compare the spread of a contagion process on these two objects, measured by the final fraction of infected nodes. In Section \ref{sect:mean_field} a mean field approximation of the model is analysed. Final discussion and further work is detailed in Section \ref{sect:disc}.

\section{Simplicial Stochastic SIR}\label{sect:simplicial_sir}
A stochastic SIR epidemic process in a simplicial complex $X$ of dimension $d$ is defined in the following way. Nodes (i.e., 0-simplices) will be either susceptible, infectious or recovered. Susceptible nodes can turn infectious, and infectious nodes eventually recover. Recovered nodes remain that way. At time 0, a randomly selected node is turned infectious, while the rest are susceptible. A $k$-simplex $\mathcal{K} \in X$ having all its nodes infected will infect a susceptible node $x$ according to a Poisson process with rate $\beta_{k+1}$ if $\mathcal{K} \cup \lbrace x \rbrace$ is a $k+1$ simplex in the complex. Each infected node remains infectious for an exponentially distributed period with mean $1/\gamma$. The process stops at the first time $T$ when zero infected nodes remain. The number of recovered nodes at time $T$ is considered the final number of infected throughout the process. Note that the regular node to node transmission observed in networks is preserved in the process definition when $k = 0$. Furthermore, the process defined as above coincides with the Markovian SIR process on networks \citep{B2020} when the simplicial complex is of dimension one (i.e., it only has nodes and edges). As far as the authors are aware, this process has not been defined in the literature. On simplicial complexes, the SIS and SIRS models have been considered \citep{Cisneros-Velarde_Bullo_2020, Iacopini_Petri_Barrat_Latora_2019, Matamalas_Gomez_Arenas_2020, wang2021simplicial}. \citet{ma2018study} study an SIR model of information transmission on a hypernetwork, but with only one transmission parameter and in discrete time. \citet{Ball_Sirl_Trapman_2009} consider an SIR process on a network with household structure, where there is a different infection rate for nodes in the same household. However, their model considers only two rates of transmission, depending on whether nodes are or not in a household. The simplicial model here defined allows for a parameter for each simplicial dimension. Similarly, \citet{Fransson_Trapman_2019} consider an SIR epidemic on a network with different transmission rates between nodes belonging to a same triangle. However, all triangles show this separate infection rate, in contrast to our model where triangles may not be included as simplices in the complex: i.e., three nodes $v_1, v_2, v_3$ may be pairwise connected in a network and thus form a triangle, but the simplex $\{v_1, v_2, v_3\}$ can be ommitted from a simplicial complex. Also, since our model has a transmission rate for each dimension, we allow groups of more than three nodes (i.e., simplicial complexes of dimension greater than 2) to have their own transmission rate, although for computational reasons, the experiments in Section \ref{sect:experiments} are limited to two-dimensional simplicial complexes. 

\begin{figure}
    \centering
    \includegraphics[width = 0.5 \linewidth]{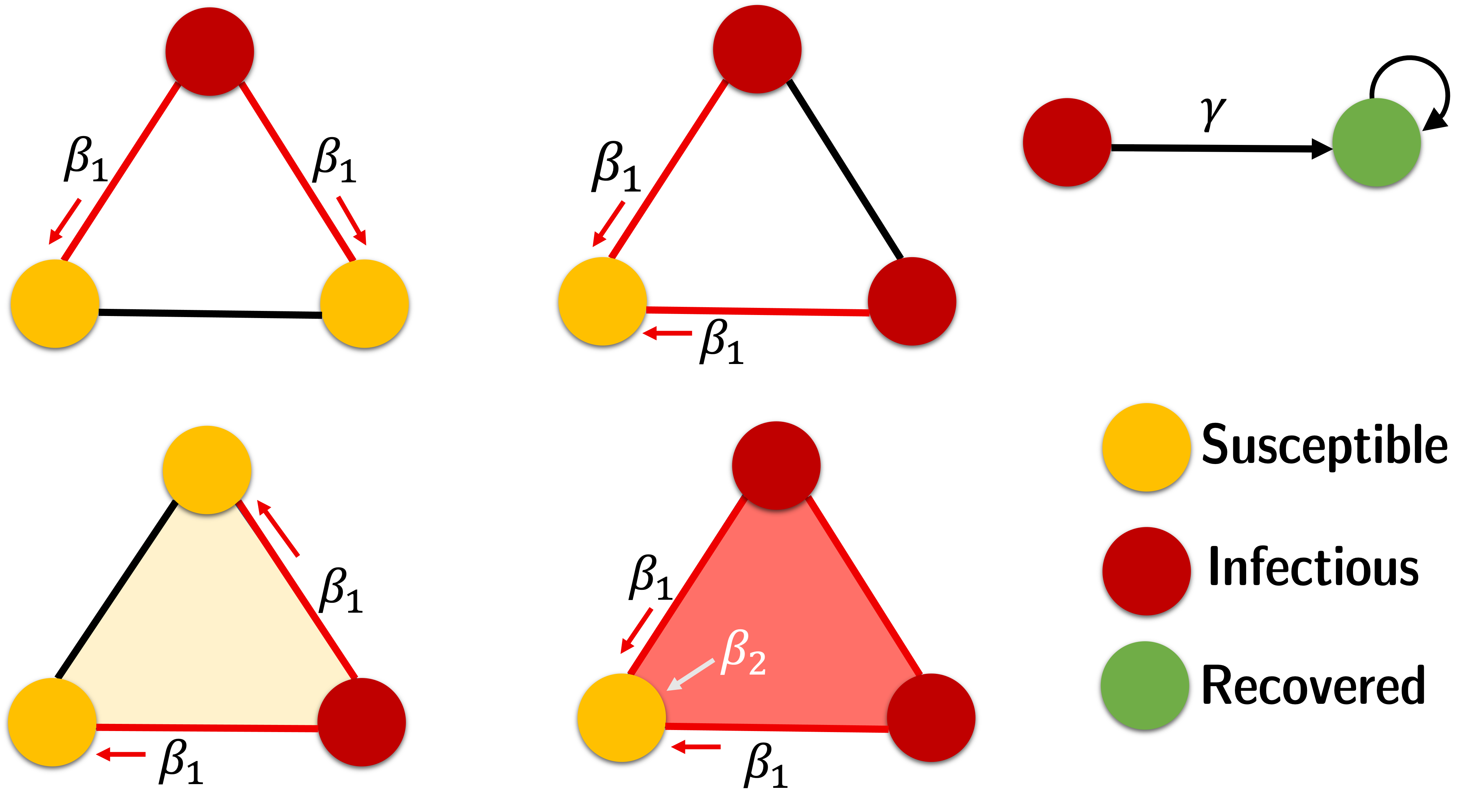}
    \caption{Representation of contagion dynamics on a simplicial complex of dimension 2.}
    \label{fig:sir_graphic}
\end{figure}

\section{Experiments}\label{sect:experiments}
\begin{figure}
		\centering
		\begin{subfigure}{0.4\linewidth}
			\includegraphics[width=\linewidth]{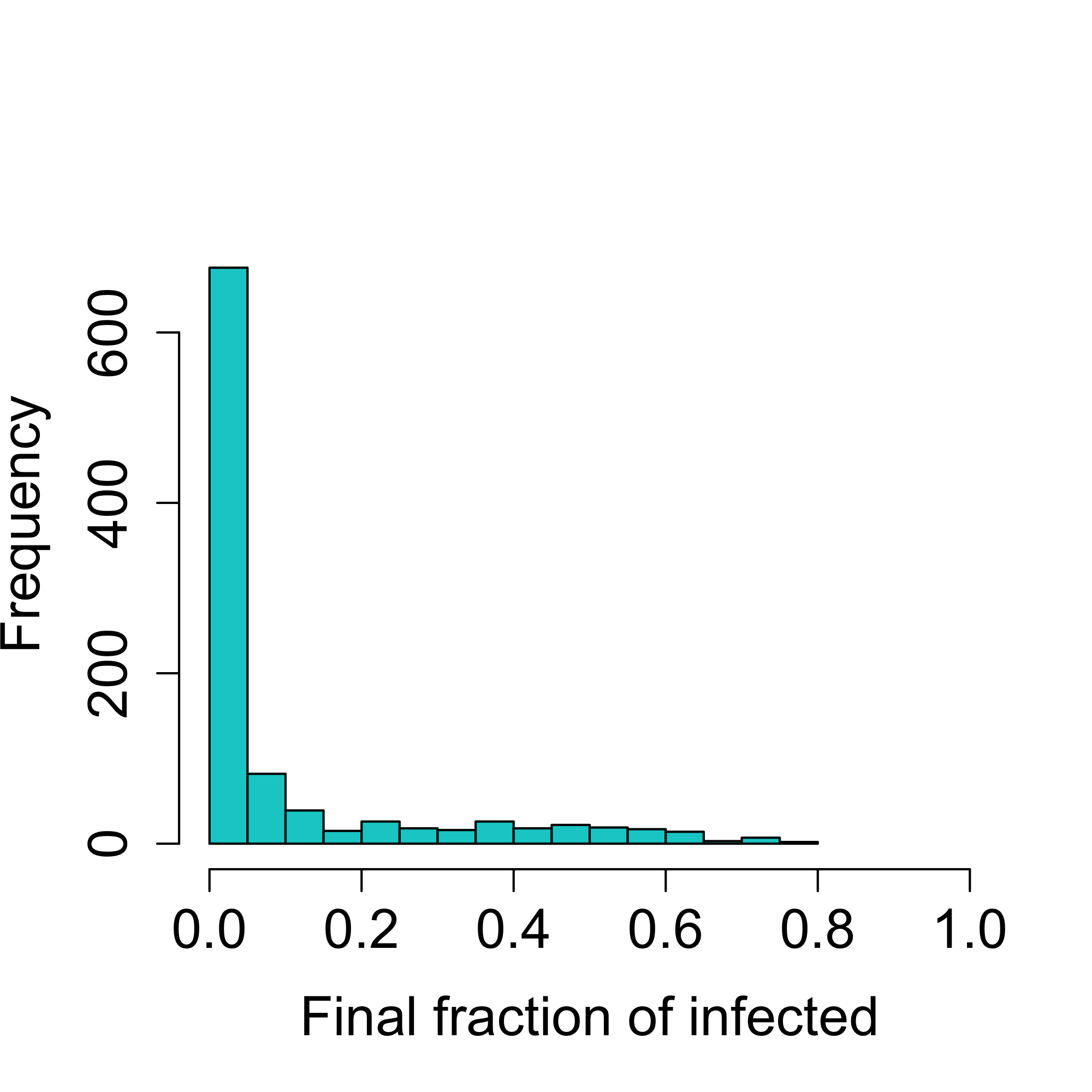}
		\end{subfigure}
		\hfill
		\begin{subfigure}{0.4\linewidth}
			\includegraphics[width=\linewidth]{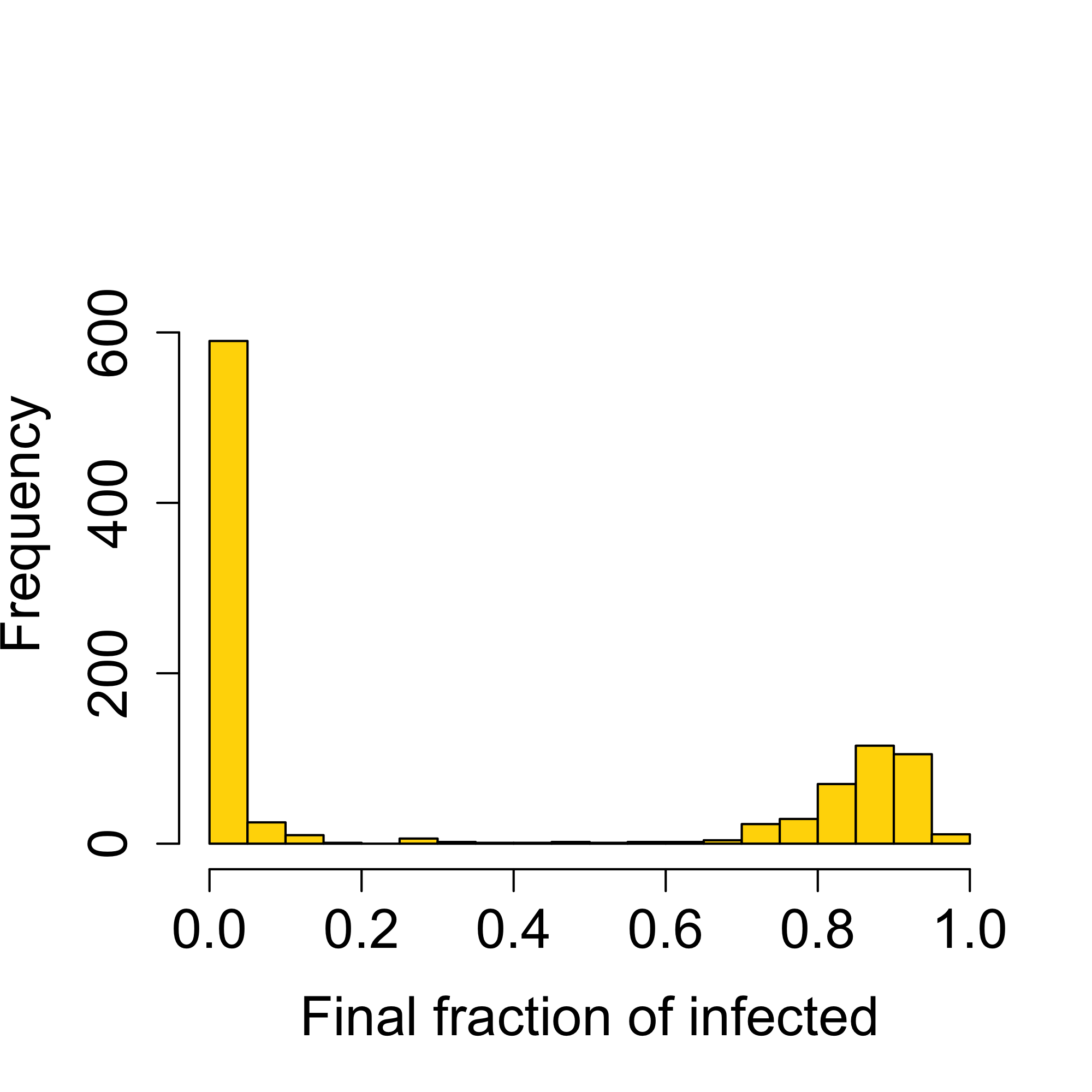}
		\end{subfigure}
		\hfill
		\begin{subfigure}{0.4\linewidth}
			\includegraphics[width=\linewidth]{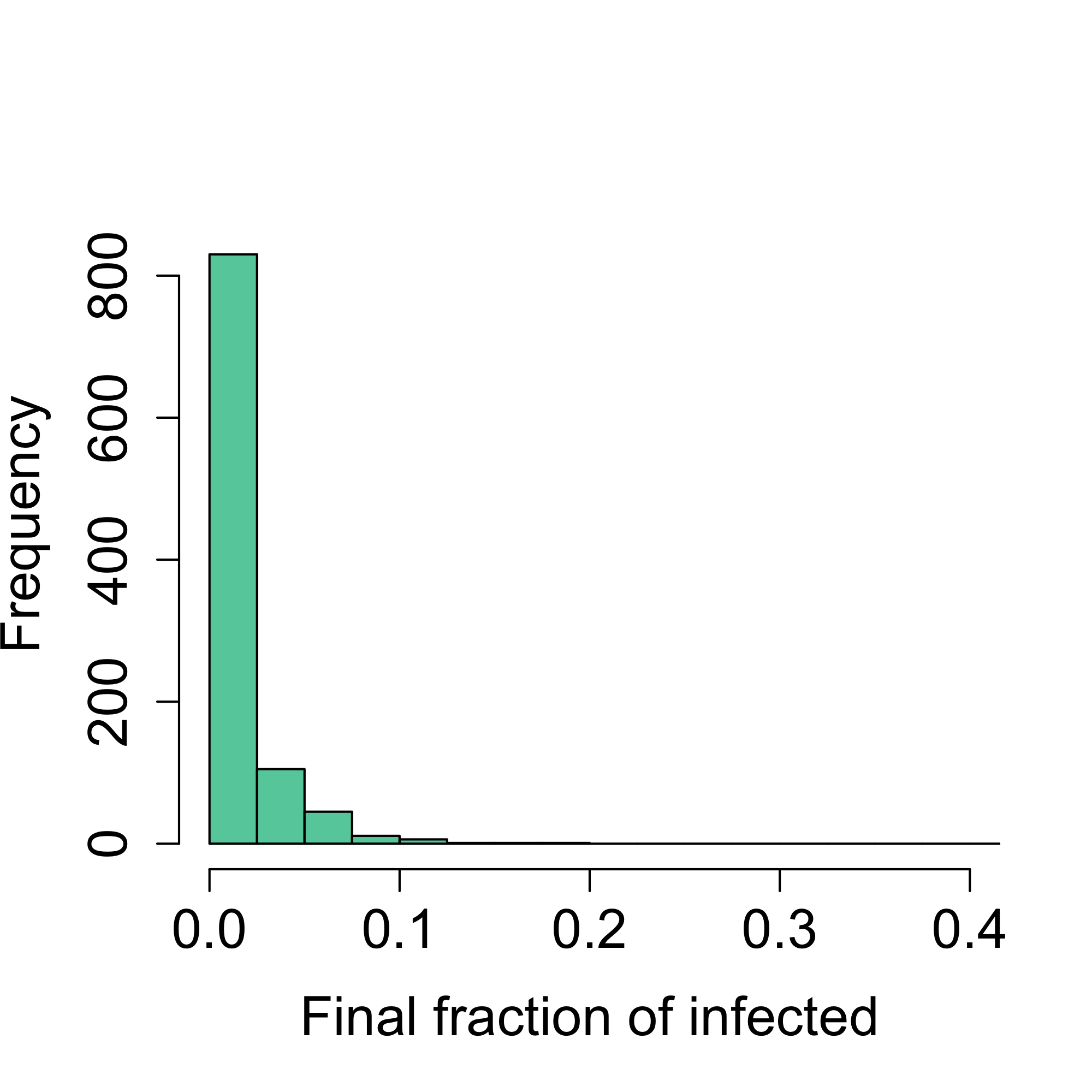}
		\end{subfigure}
		\hfill 
		\begin{subfigure}{0.4\linewidth}
			\includegraphics[width=\linewidth]{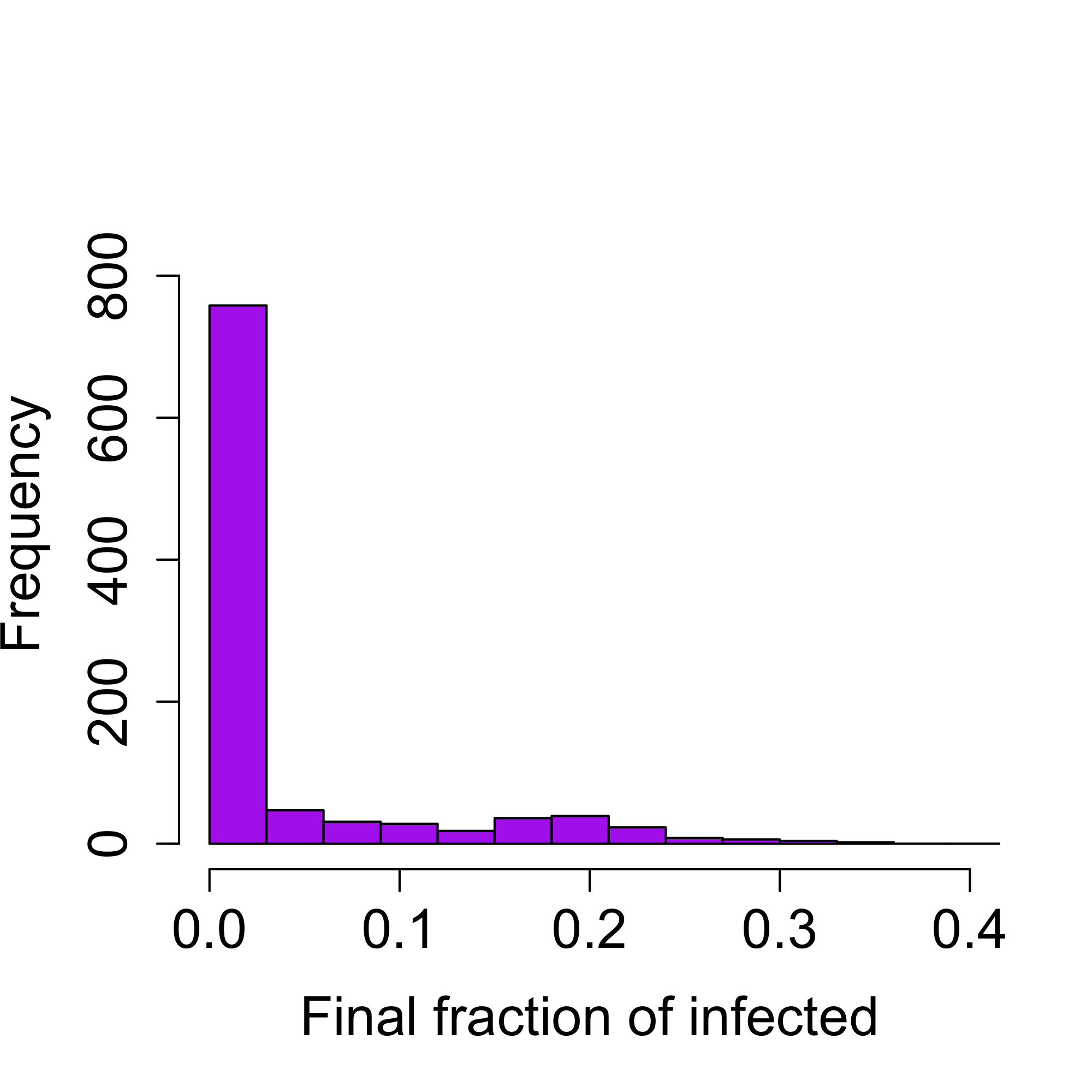}
		\end{subfigure}
	\caption{Histograms for the final fraction of infected for the simplicial stochastic SIR on graphs (left) and their two-dimensional clique complexes (right). Graphs used were an Erd\H{o}s-Rényi graph (top) and a high school social network (bottom). Transmission rates were $\beta_1 = 1$ in all cases, $\beta_2 = 0$ for graphs and $\beta_2 = 15$ for the corresponding clique complexes. Recovery rate was $\gamma = 9$ for all cases}
	\label{fig:histograms}
\end{figure}

As a first step to determine whether the simplicial higher order interactions have an effect on the spread of an SIR epidemic, simulations of the process are performed for an Erd\H{o}s-Rényi graph $G(n = 120, p = 0.1)$ and its two-dimensional clique complex, i.e., all triangles of the graph are included as two-simplices. The resulting network has 120 nodes, 734 edges, 288 cliques of size 3, average degree of $\langle k_1 \rangle =  12.23$, each node belonging to an average of $\langle k_2 \rangle = 7.2$ two-simplices. The distribution of the final fraction of infected nodes in both scenarios will be studied.  The parameters used are transmission rates $\beta_1 = 1, \beta_2 = 0$ in the case of the graph (one-dimensional complex) and $\beta_1 = 1, \beta_2 = 15$ in the case of its two-dimensional clique complex. In both cases, the recovery rate is $\gamma = 9$. 
A thousand runs are done for each process. The top histograms in Figure \ref{fig:histograms} show the distributions of the final fraction of infected nodes, which can be compared for the process with $\beta_2 = 0$ (left, epidemic on a network) and the process with $\beta_2 = 15$ (right, epidemic on the two-dimensional clique complex). The difference in frequency with which the epidemic takes off is noticeable.

The same process is considered in a network of contacts between students in a high school in Marseilles, France \citep{MFBV2015} and its two-dimensional clique complex. The network has 120 nodes, 348 edges, 272 cliques of size 3, $\langle k_1 \rangle =  5.8$ and $\langle k_2 \rangle = 6.8$.  Simulations with parameters $\beta_1 = 1, \beta_2 = 0, \gamma = 9$ and $\beta_1 = 1, \beta_2 = 15, \gamma = 9$ are done with a thousand runs each. Similarly, the distribution of the final fraction of infected is shown in the bottom histograms of Figure \ref{fig:histograms}. It is observed how introducing the simplicial contagion leads to a larger frequency of outbreaks. 

\begin{figure}
		\centering
		\begin{subfigure}{0.4\linewidth}
			\includegraphics[width=\linewidth]{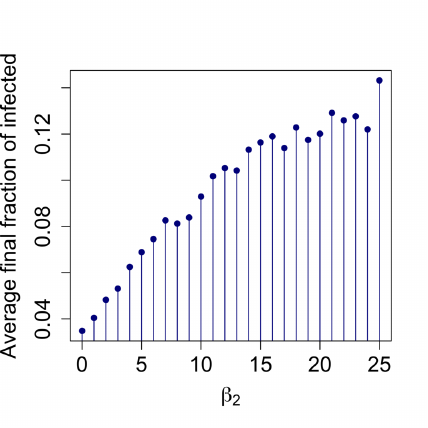}
		\end{subfigure}
		\hfill
		\begin{subfigure}{0.43\linewidth}
			\includegraphics[width=\linewidth]{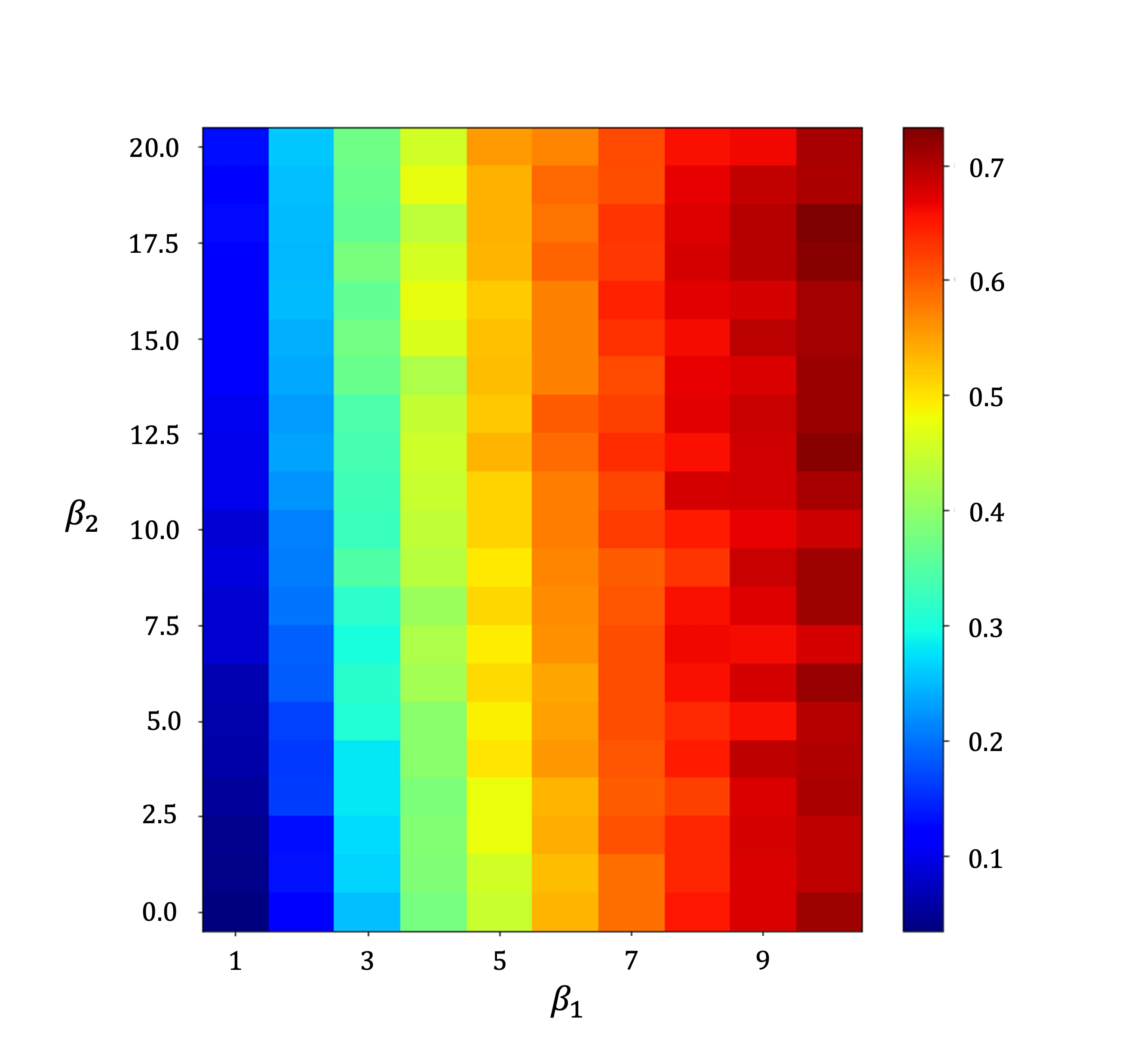}
		\end{subfigure}
		\\ \vspace{0.5 cm}
		\caption{Simplicial complex used consisted of 120 nodes, $\langle k_1 \rangle = 5.8, \langle k_2 \rangle = 6.8$. \textbf{Left:} Average final fraction of infected over a thousand runs for fixed $\beta_1 = 1, \gamma = 5$ and different values of $\beta_2$. \textbf{Right:} Average final fraction of infected over a thousand runs for fixed $\gamma = 5$ and different values of $\beta_1, \beta_2$.}
		\label{fig:variations}
\end{figure}

 The use of $\beta_2 = 15$ in both cases was rather arbitrary. To further support our findings, the \textit{average} final fraction of infected after 1000 runs of the epidemic process on the contact network were done for fixed $\beta_1 = 1, \gamma = 5$ and varying $\beta_2 = 0, 1, \dots, 25$, and for fixed $\gamma = 5$ and varying $\beta_1 = 1, 2, \dots, 10, \beta_2 = 0, 0.5, 1, \dots, 20$. The results are displayed in Figure \ref{fig:variations}. In the case of the heat map in Figure \ref{fig:variations}, if the parameter $\beta_2$ had no effect, one would expect to see an increment on the average fraction of infected only when $\beta_1$ increases, which is the left to right direction in the heatmap. However, a bottom-to-top increment is seen when fixing any of the $\beta_1$ values, corresponding to an increment of the parameter $\beta_2$. This effect is less noticeable for larger $\beta_1$ values. 




\section{Mean field approach}\label{sect:mean_field}
A mean-field approximation, under an homogeneous degree hypothesis is considered. For a simplicial complex of dimension $D$, transmission rates $\beta_1, \dots, \beta_D$, recovery rate $\gamma$ and average degree $\langle k_d \rangle$ for the $d$-dimensional simplices, the mean field equations are Equations \ref{eq:mf_s} -- \ref{eq:mf_r}:
\begin{align}
    s'(t) &= - \sum_{d = 1}^D \beta_d \langle k_d \rangle i(t)^d s(t) \label{eq:mf_s},\\
    i'(t) &= - \gamma i(t) + \sum_{d = 1}^D \beta_d \langle k_d \rangle i(t)^d s(t), \label{eq:mf_i},\\
    r'(t) &= \gamma i(t) \label{eq:mf_r}.
\end{align}
At any given time $s + i + r = 1$. The initial conditions satisfy, $s_o + i_o + r_o = 1$, $r_o = 0$.
Due to the dissipative dynamics imposed by the recovered rate equation,  
the epidemic has an absorbing state, $s_f + i_f + r_f = 1$, $i_f = 0$.
Of particular interest is the total fraction of the total population of $N$
nodes which gets 
infected during the entire
epidemic, $I \equiv \int _{0}^{\infty} i(t) dt$.
Clearly, the total number of infected nodes is
$NI$ and 
$r_f = I $.

Assuming that $i$, $s$ and $r$ are analytic functions,
the mean field equations can be used to 
study the system close to any given point in the state space $(s_a, i_a, r_a)$ by taking
 $i(t) = i_a + \phi  (t)$, $s(t) = s_a + \sigma  (t)$
and $r(t) = r_a + \rho  (t)$, where $\phi$, $\sigma$ and 
$\rho$ are small quantities.
Substitution in the susceptible rate equation gives,
\begin{align}
     \sigma '(t) &= - [s_a + \sigma  (t)] \sum_{d = 1}^D \beta_d \langle k_d \rangle [i_a + \phi  (t)]^d \\ &=  \nonumber
       - [s_a + \sigma  (t)] 
      \sum_{d = 1}^D \beta_d \langle k_d \rangle 
     \left [ \sum_{r=0} ^{d} i^{d-r}_{a} \phi^{r}  (t)
     \binom{d}{r}  \right].
\end{align}
By the corresponding substitution in the infected rate equation, 
the system 
is expressed by the following two coupled differential equations,
\begin{eqnarray}\label{expansion}
     \sigma '(t) + 
      [s_a + \sigma  (t)]\left [ 
     \sum_{d = 1}^D \beta_d \langle k_d \rangle \left[ i_a^{d} +
       i^{d-1}_{a} \phi  (t)
      \frac{d!}{(d-1)!} \right ] \right] \\ \nonumber
      = 
      -  [s_a + \sigma  (t)] \sum_{d = 1}^D \beta_d \langle k_d \rangle 
     \left [\sum_{r=2} ^{d} i^{d-r}_{a} \phi^{r}  (t)
     \binom{d}{k}  \right], \\ \nonumber
\end{eqnarray} 
\begin{eqnarray}\nonumber
     \phi '(t) +
     \gamma \phi  (t) 
     - [s_a + \sigma  (t)] \left [
     \sum_{d = 1}^D \beta_d \langle k_d \rangle 
     \left [ i_a^{d} + i^{d-1}_{a} \phi  (t)
     \frac{d!}{(d-1)!} \right ] \right] \\ \nonumber
     = 
      [s_a + \sigma  (t)] \sum_{d = 1}^D \beta_d \langle k_d \rangle 
     \left [\sum_{r=2} ^{d} i^{d-r}_{a} \phi^{r}  (t)
     \binom{d}{k}  \right ] .
\end{eqnarray}
Up to first order in $\phi$, Equation \ref{expansion} reads,
\begin{eqnarray}\label{expansion2}
     \sigma '(t) 
     + [s_a + \sigma  (t)]\left [  
     \sum_{d = 1}^D \beta_d \langle k_d \rangle 
      \left [ i_a^{d} + d i_a^{d-1} \phi  (t)
       \right ] \right]
      = 
     0
\\ \nonumber
     \phi '(t) 
     +  \gamma \phi  (t) 
     - [s_a + \sigma  (t)] \left [ 
     \sum_{d = 1}^D \beta_d \langle k_d \rangle 
     \left [ i_a^{d} + d i_a^{d-1} \phi (t)
      \right ] \right] 
     = 0
\end{eqnarray}
The fact that the only possible value for $i_f$
is zero implies that 
the absorbing state 
tends to a dynamic system that 
is independent of the
simplices of order greater than $D=1$. To see this, note
that close to $i_f = 0$, only the simplex $D=1$
survives in Equation \ref{expansion2},
\begin{eqnarray}
   s ' \to -\beta_1 \langle k_1 \rangle s(t) \phi (t) \\ \nonumber
   r' \to \gamma \phi (t),
\end{eqnarray}
so
\begin{eqnarray}
    \frac{ds}{dr} = \frac{-\beta_1 \langle k_1 \rangle s(t)}{\gamma}.
\end{eqnarray}
Therefore,
\begin{eqnarray} \label{absorb1}
    s_f = s_\tau e ^{ \frac{\beta_1 \langle k_1 \rangle}{\gamma} (1 - s_f) }  , 
\end{eqnarray}
where $\tau$ is a characteristic time scale.

\begin{figure}
		\centering
		\begin{subfigure}{0.47\linewidth}
			\includegraphics[width=\linewidth]{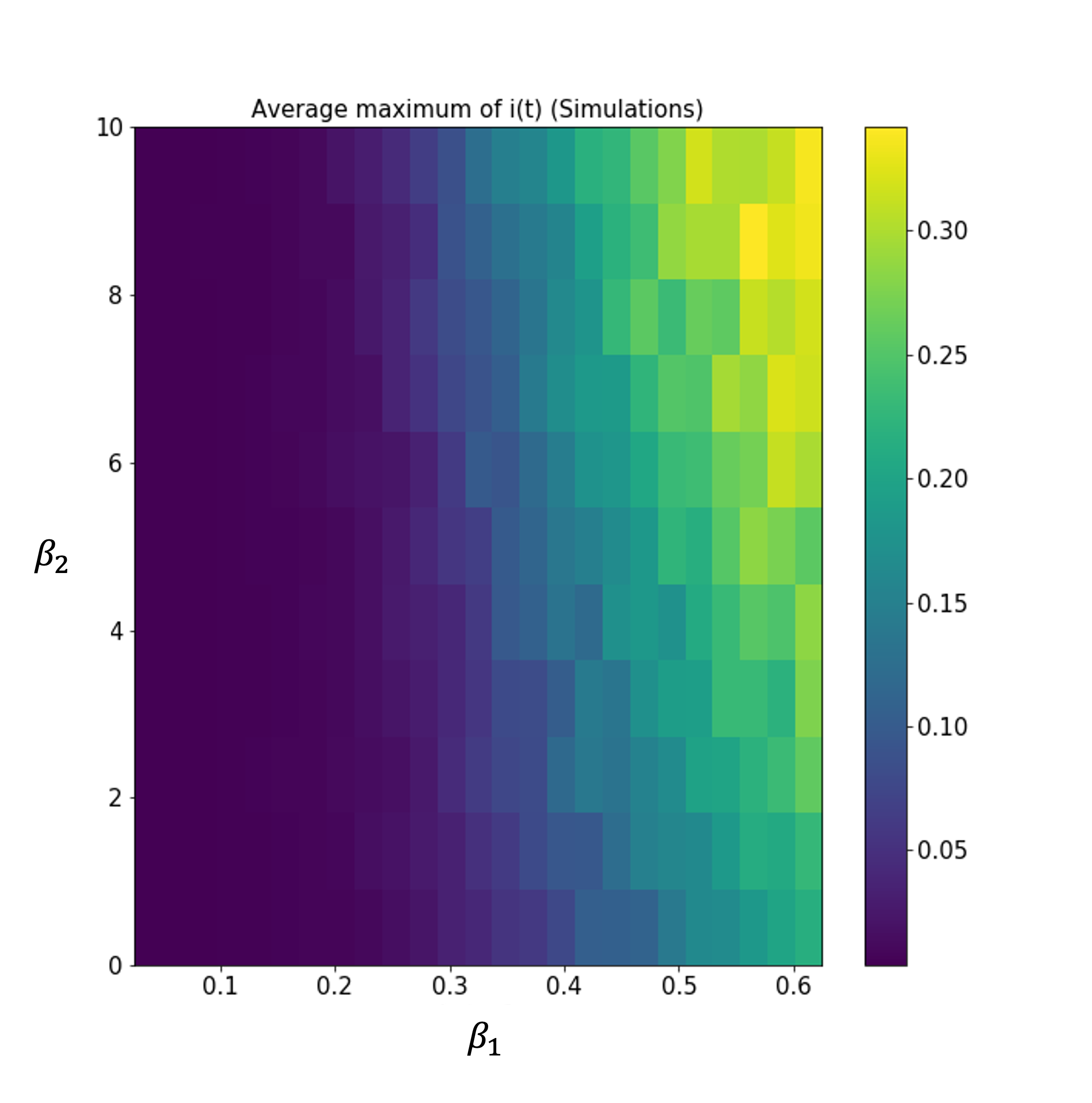}
		\end{subfigure}	
		\hfill
		\begin{subfigure}{0.47\linewidth}
			\includegraphics[width=\linewidth]{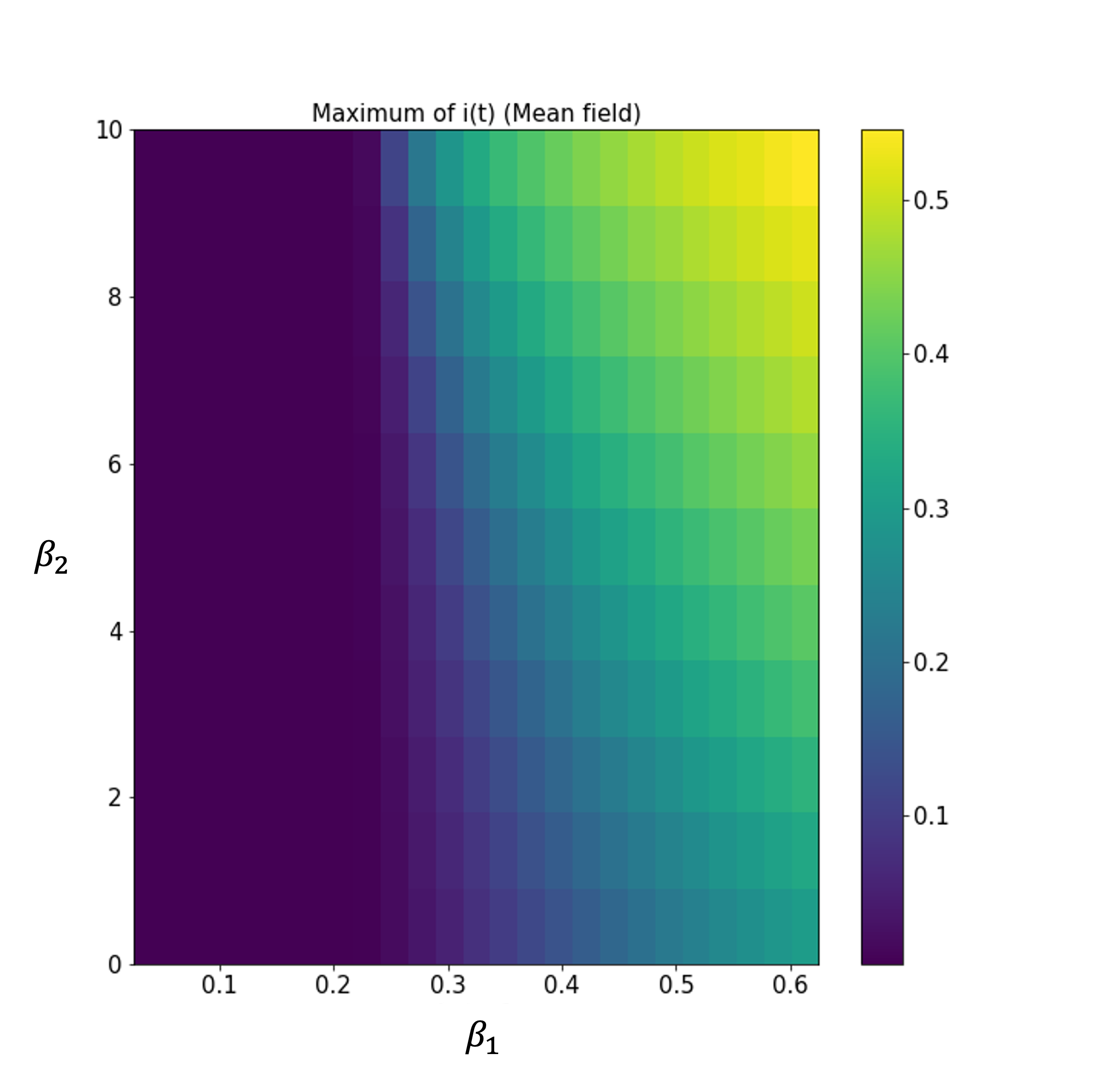}
		\end{subfigure}	
		\begin{subfigure}{0.47\linewidth}
			\includegraphics[width=\linewidth]{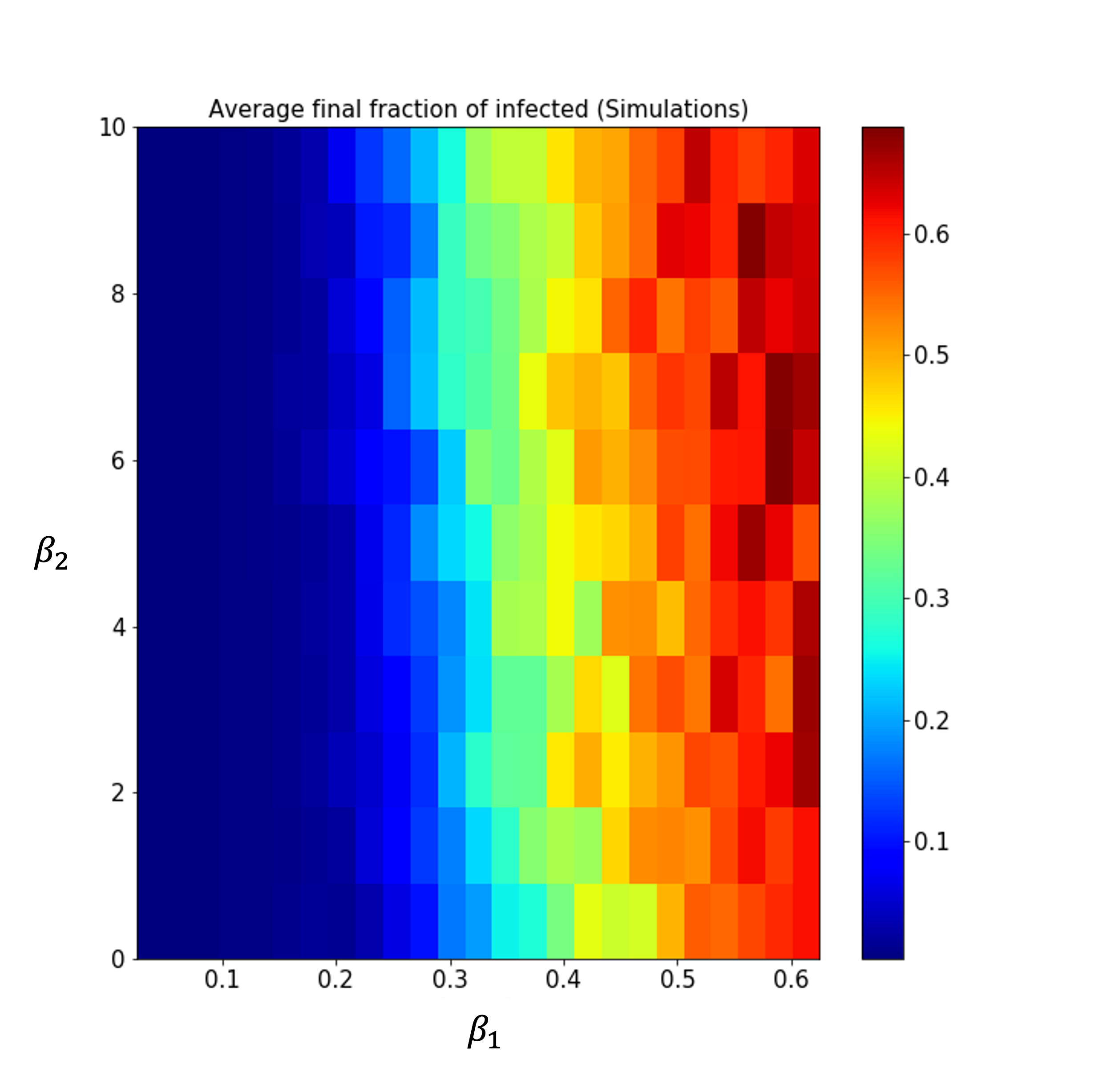}
		\end{subfigure}	
		\hfill
		\begin{subfigure}{0.47\linewidth}
			\includegraphics[width=\linewidth]{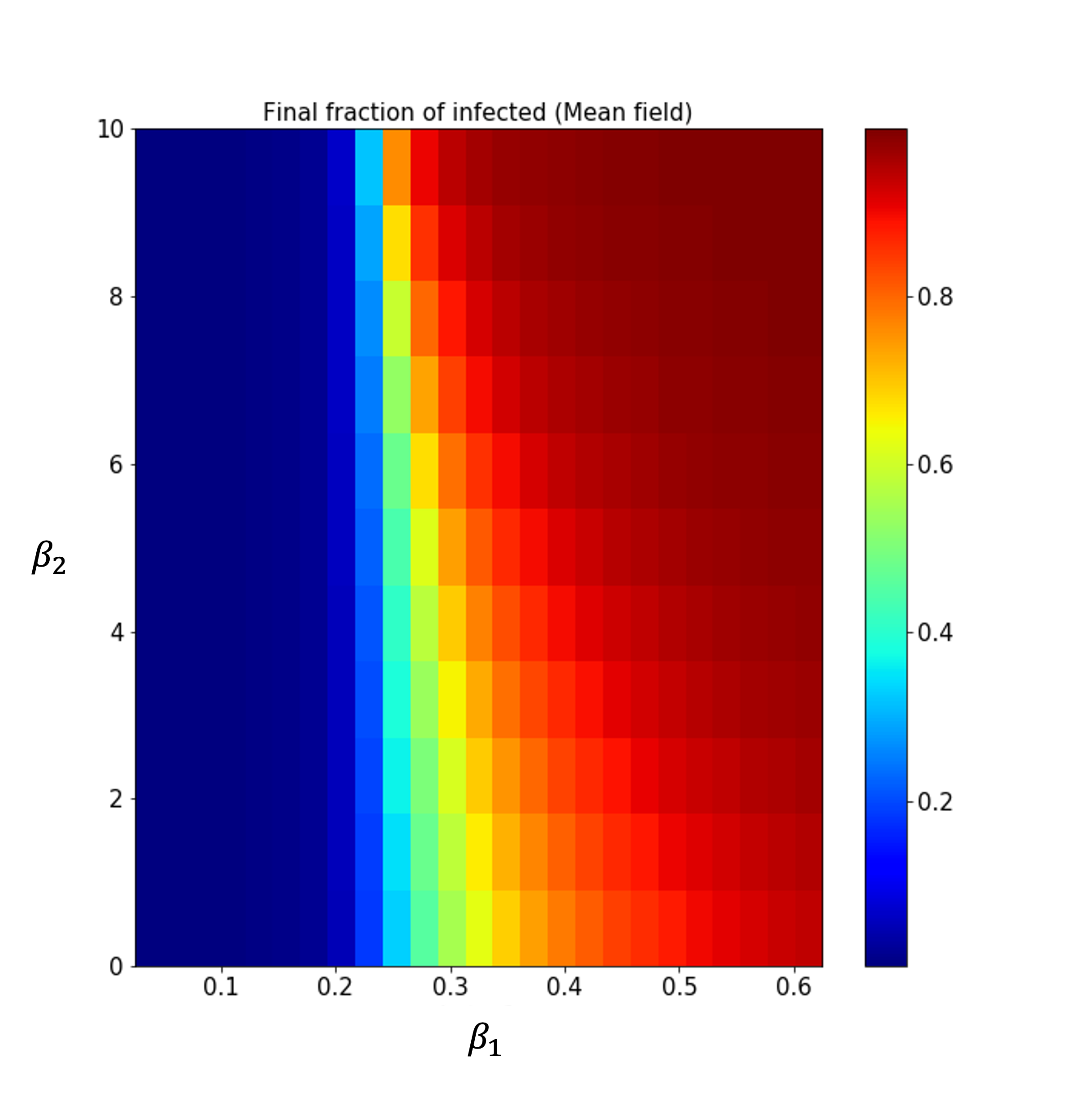}
		\end{subfigure}
		\caption{Heatmaps comparing average final of infected and maximum of infected obtained by simulation vs. mean field approximation on a random simplicial complex with 300 nodes, $\langle k_1 \rangle \approx 50$, $\langle k_2 \rangle \approx 10$, $\gamma = 10$.}
		\label{fig:heatmaps_sim_vs_mf}
\end{figure}

\begin{figure}
		\centering
		\begin{subfigure}{0.47\linewidth}
			\includegraphics[width=\linewidth]{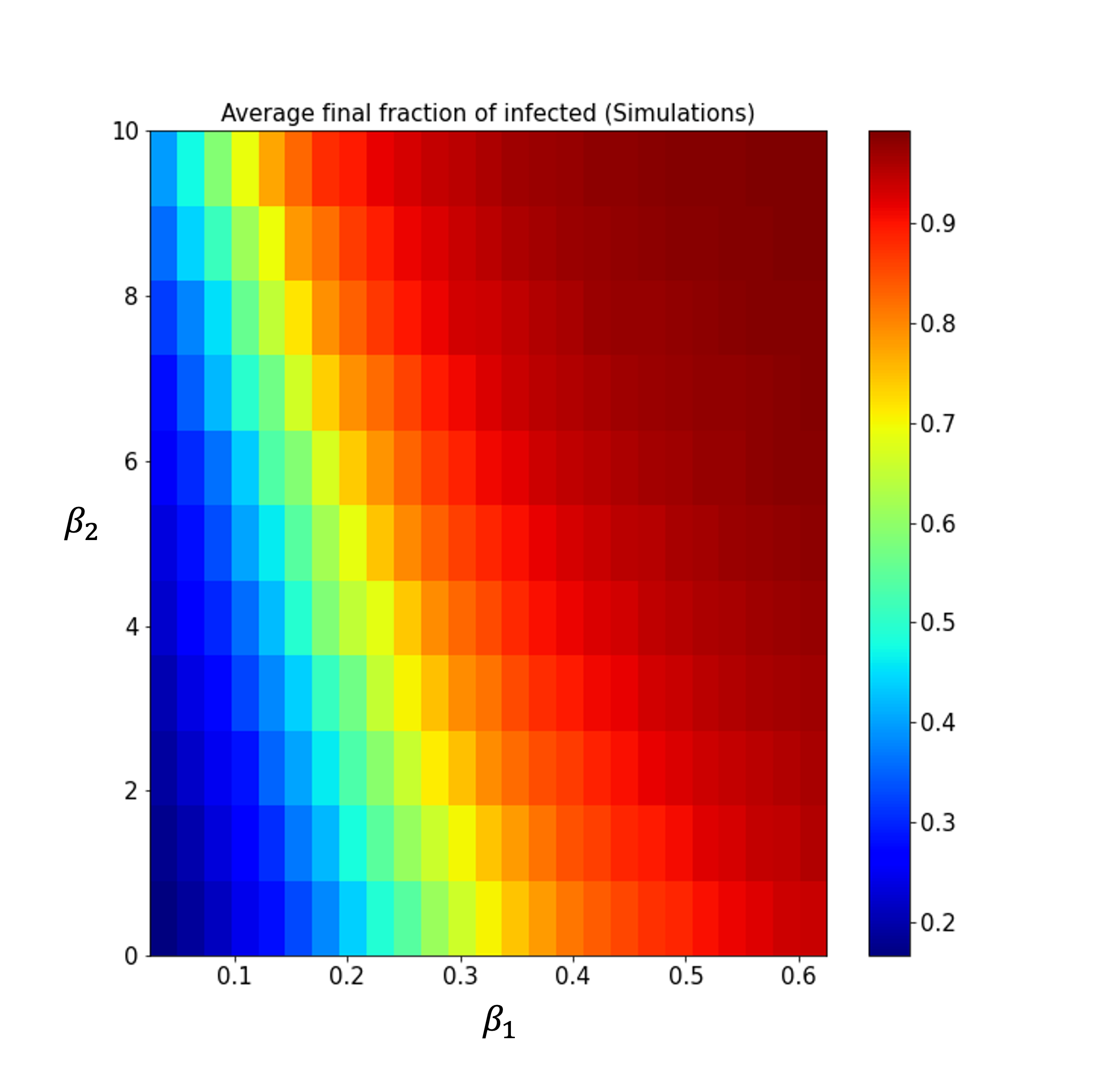}
		\end{subfigure}	
		\hfill
		\begin{subfigure}{0.47\linewidth}
			\includegraphics[width=\linewidth]{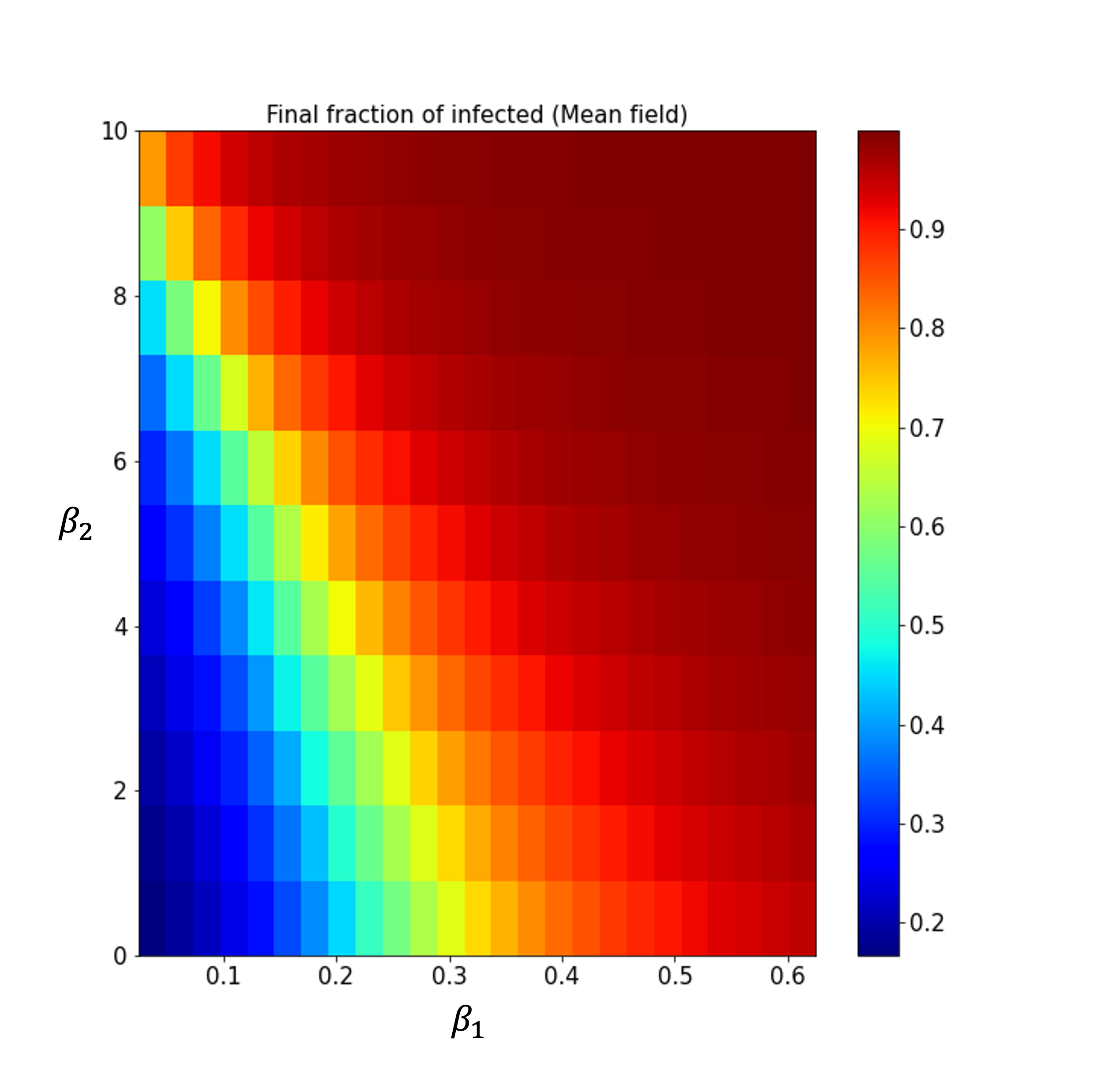}
		\end{subfigure}
		\caption{Heatmaps comparing average final of infected obtained by simulation vs. mean field approximation on a random simplicial complex with 300 nodes, $\langle k_1 \rangle \approx 50$, $\langle k_2 \rangle \approx 10$, $\gamma = 10$ but considering 45 (15\%) of initially infected nodes.}
		\label{fig:heatmaps_sim_vs_mf_ini_inf}
\end{figure}

Under the additional assumption that $|\sigma| < |\phi|$, which is 
reasonable given that the recovery rate must be lesser than the 
infection rate in order to have an outbreak at any arbitrary initial susceptible 
population size, the equation for $\phi$
\begin{eqnarray}
     \phi '(t) 
     +  \phi  (t) \left[ \gamma  - [s_a + \sigma  (t)]
     \sum_{d = 1}^D \beta_d \langle k_d \rangle  i_a^{d-1} d
       \right] - [s_a + \sigma  (t)] \sum_{d = 1}^D \beta_d \langle k_d \rangle  i_a^{d} 
     = 0
\end{eqnarray}
has the solution
\begin{eqnarray} \label{solphi}
   \phi  (t) = \frac{s_a \sum_{d = 1}^D \beta_d \langle k_d \rangle  i_a^{d}}
       {\left[ \gamma - 
      s_a  \sum_{d = 1}^D \beta_d \langle k_d \rangle   i_a^{d-1} d \right]} 
      \left[ 1 - e ^{- \left[ \gamma - 
      s_a  \sum_{d = 1}^D \beta_d \langle k_d \rangle   i_a^{d-1} d \right] t } \right].
\end{eqnarray}
From Eq. (\ref{solphi}) it follows the outbreak threshold for the 
epidemics, which is affected by the higher order interactions of
simplices $D>1$, is
\begin{eqnarray}
    s_o \sum_{d = 1}^D \beta_d \langle k_d \rangle i_o^{d-1} d
      > \gamma .
\end{eqnarray}
Clearly, the well known outbreak condition for a SIR epidemics
in a graph $D=1$ is recovered,
$ s_o > \frac {\gamma}{\beta_1 \langle k_1 \rangle}$.
Furthermore, the threshold condition can be verified with the numerical solution of the mean field equations. A process with $\gamma = 10$, $\langle k_1 \rangle \approx 50$, $\langle k_2 \rangle \approx 10$, and one initially infected node out of 300, would have a threshold for values above the line $\beta_2 = 45000/299 - 750 \beta_1$ in a $(\beta_1, \beta_2)$ plane, which is the straight line which intersects at $\beta_1 = 60/299 \approx 0.2$ and $\beta_2 = 45000/299 \approx 150$, which can be observed in Figure \ref{fig:heatmaps_sim_vs_mf} as the final fraction of infected in the mean field case (bottom right) is close to zero for this region. The small magnitude of initially infected nodes makes the effect of $\beta_2$ small for this threshold condition. If, however, an initial fraction of 15\% infected nodes (45/300) is considered, the threshold region is that above the line $\beta_2 = 200/51 - (50/3)\beta_1$ in a $(\beta_1,  \beta_2)$ plane, which is the straight line which intersects at $\beta_1 = 4/14 \approx 0.235$ and $\beta_2 = 200/51 \approx 3.92$. This can be observed in Figure \ref{fig:heatmaps_sim_vs_mf_ini_inf}.

From Eq. (\ref{solphi}) the characteristic time scale for the epidemic $\tau$, 
can be defined as
\begin{eqnarray}\label{tau}
    \tau = \frac{1}{\left[ \gamma - 
      s_a  \sum_{d = 1}^D \beta_d \langle k_d \rangle   i_a^{d-1} d \right]}.
\end{eqnarray}
At time $\tau$, an epidemic outbreak in 
a susceptible population essentially starts to die out, $\phi \to 0$. 
It follows that exactly in $t = \tau$, 
\begin{align}
    0 &= - \gamma i(\tau) + 
    \sum_{d = 1}^D \beta_d \langle k_d \rangle i(\tau)^d s(\tau).
\end{align}
By the equation for $s'$ and
introducing the definitions $i_\tau \equiv i(\tau)$ and $s_\tau \equiv s(\tau)$,
\begin{eqnarray}\label{tau2}
   s_\tau = s_o e^{-\tau  i_\tau
\left [ \beta_1 \langle k_1 \rangle + 
2 \beta_2 \langle k_2 \rangle  i_\tau + ... 
+ D \beta_D \langle k_D \rangle i_\tau^{D-1} \right  ] }, \\ \nonumber
\beta_1 \langle k_1 \rangle + 2 \beta_2 \langle k_2 \rangle  i_\tau
   + 3 \beta_3 \langle k_3 \rangle i_\tau^{2}+...
   + D \beta_D \langle k_D \rangle i_\tau^{D-1}
      = \frac{\gamma}{s_{\tau}}.
\end{eqnarray}
By the definition of $\tau$, the epidemic peak is bounded from below 
by $i_\tau$. For $D=2$,
\begin{eqnarray}\label{bound}
   i_\tau = \frac{1}{2\beta_2 \langle k_2 \rangle}
    \left [ \frac{\gamma}{s_\tau} - \beta_1 \langle k_1 \rangle \right   ]
    \leq i_{*}.
\end{eqnarray}
Therefore for $D=2$ there is a bound for the value $s$ at the infection peak, 
denoted by $s_*$, 
\begin{eqnarray}\label{bound2}
      s_* \leq \frac{\gamma}{\beta_1 \langle k_1 \rangle}.
\end{eqnarray}

The mean field approximation is tested numerically by comparing its results with those of simulations in random simplicial complexes.
In the Figure \ref{fig:variations} the final fraction of infected nodes $I$,
given by averaging over $1000$ simulations of the microscopic
stochastic process is reported. The graph at the right
of the Figure \ref{fig:variations} shows a parametric sweep over the range
$\beta_1 \in [1, 10]$, $\beta_2 \in [0, 20]$ with $\gamma = 5$ fixed.  
The graph in the left is the slice at $\beta_1 = 1$. 
From Equation \ref{bound} and taking into account 
that $s_f \leq s_\tau$, $I \geq i_*$ and $s_f = 1 - I$,
mean field predicts 
the bound given by the relation,
\begin{eqnarray}\label{bound3}
    I \geq 1 - \frac{\gamma}{\beta_1 \langle k_1 \rangle} .
\end{eqnarray}
This bound is experimentally verified for
the parametric sweep considered in the microscopic simulations
reported in
Figure \ref{fig:variations}
and by the simulations and numerical solutions of the mean field equations 
reported in the Figures \ref{fig:heatmaps_sim_vs_mf} and 
\ref{fig:heatmaps_sim_vs_mf_ini_inf}.
For instance, in the conditions at the left
graph of Figure \ref{fig:variations}, the resulting bound is
$I \geq 0.1379$.
The conditions considered in the 
Figures \ref{fig:heatmaps_sim_vs_mf} and 
\ref{fig:heatmaps_sim_vs_mf_ini_inf} on the other hand, obey the 
bound $I \geq 0.6799$.
Both analytical bounds are 
in complete accordance with observations.
The interplay between fluctuations and high order
interactions is however not completely captured by the 
homogeneous mean field theory, which is 
exemplified by Figure \ref{fig:histograms}, where the
histograms of the values for $I$ sampled from the 
simulations are reported. The uncertainty in
$I$ is clearly influenced by the simplicial 
interactions.

\begin{figure}
		\centering
		\begin{subfigure}{0.47\linewidth}
			\includegraphics[width=\linewidth]{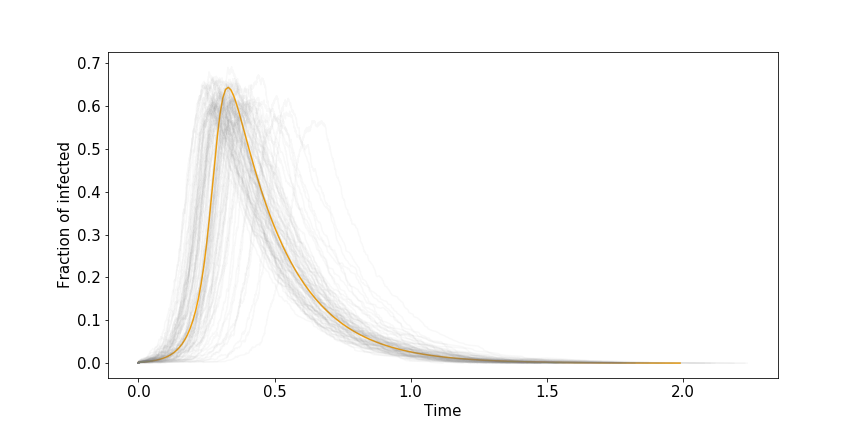}
		\end{subfigure}
		\hfill
		\begin{subfigure}{0.47\linewidth}
			\includegraphics[width=\linewidth]{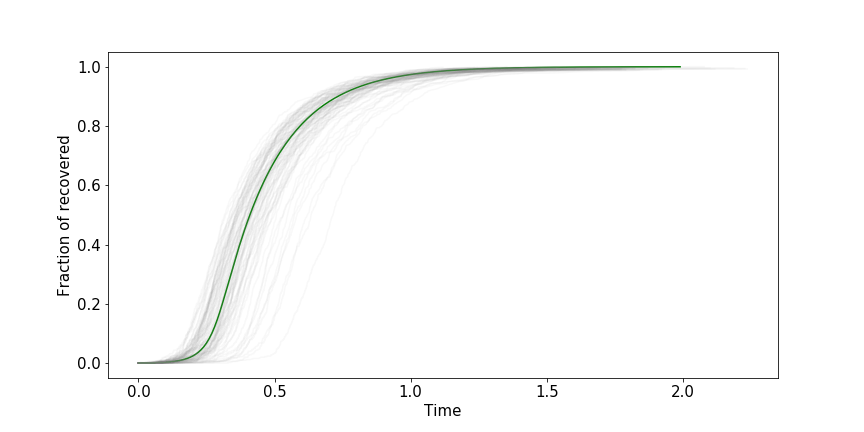}
		\end{subfigure}
		\caption{Comparing the evolution of infectious (left) and recovered (right) nodes in the realizations of a hundred simulations (gray) and the numerical solution of the mean field equations (color). Simplicial complex used had 500 nodes, $\langle k_1 \rangle \approx 25, \langle k_2 \rangle \approx 10$, and the parameters were $\beta_1 = 1, \beta_2 = 5, \gamma = 5$. }
		\label{fig:realizations_sim_vs_mf}
\end{figure}

 Figure \ref{fig:realizations_sim_vs_mf} shows the evolution of infectious and recovered individuals in time on a simplicial complex with 500 nodes, $\langle k_1 \rangle \approx 25, \langle k_2 \rangle \approx 10$ and an epidemic with parameters $\beta_1 = 1, \beta_2 = 5, \gamma = 5$. 
Notice that the bound (\ref{bound}) is consistent with the observed peak.

In Figure \ref{fig:heatmaps_sim_vs_mf} (top), the maximum of infected nodes during the epidemic is compared between mean-field approximations and simulations (in this latter case, via an average of two hundred repetitions). Similarly, the final fraction of infected is compared in Figure \ref{fig:heatmaps_sim_vs_mf} (bottom). All the comparisons were done on a simplicial complex with 300 nodes, $\langle k_1 \rangle \approx 50$, $\langle k_2 \rangle \approx 10$, and epidemics with parameters $\beta_1 = 0.025, 0.05, \dots, 0.625$, $\beta_2 = 0, 1, \dots, 10$, $\gamma = 10$. It is seen that our mean field approximation is consistent with our simulation results, which allows the possibility of future work to be done without expensive simulations.

\section{Discussion}\label{sect:disc}
The dissipative dynamics of the SIR contagion process permits a 
very complete mean field characterization by which it is even possible to obtain
an explicit solution for the changes in the fraction of infected nodes close 
to any given state space point for any simplex dimension. 
Other similar contagion processes like SIS
\citep{Cisneros-Velarde_Bullo_2020, Iacopini_Petri_Barrat_Latora_2019, Matamalas_Gomez_Arenas_2020} or SIRS \citep{wang2021simplicial}
have been recently studied on simplicial complexes, but our model offers a framework
in which the mean field theory, 
besides its good agreement with the underlying microscopic process 
(as shown in Figures \ref{fig:realizations_sim_vs_mf}, 
\ref{fig:heatmaps_sim_vs_mf} and \ref{fig:heatmaps_sim_vs_mf_ini_inf}),
it is also capable of describing the effects of arbitrarily higher order
interactions. This makes our model an ideal framework for
the study of variations beyond contagion processes on simplicial networks.
From the results in Sections \ref{sect:experiments} and \ref{sect:mean_field}
it is clear that the presence of higher order simplicials has an impact
in all of the prominent features of the epidemic. Take for instance
the epidemic's peak, which is bounded according to Eq. (\ref{bound})
at interactions up to $D=2$. It turns out that
an increase in the strength of the
interactions within the simplex has the not very intuitive 
effect of diminishing the lower bound for the infection's peak.
The following intriguing remarks are however worth noting at this 
matters. It is plausible to consider the simplicial interactions
to model other types of relationships among nodes besides infection,
for example risk perception or social pressure-induced prophylaxis. 
In these situations the interaction strength parameters very well can 
be of positive or negative sign. 
In such a case,
the mathematical structure 
of the mean field linearized solutions does not forbid
the existence of stationary states different from the 
absorbing $i=0$ state. A second remark is that
the microscopic simulations shown in Figure \ref{fig:histograms}
indicate that although the average final number of
infected nodes is almost independent of the simplicial interactions, 
the fluctuations around that average are not and this
can be indicative of some interplay
between the simplicial interactions and the 
underlying stochasticity.
This aspect cannot be analyzed by the developed homogeneous 
mean field description, so further study via more extensive 
simulations or more advanced mean field approximations
is worth to pursue for future works on the model. 

\subsection{Concluding remarks}
Overall, simplicial complexes look like a promising tool to generalize many-bodied interactions in complex networks. The present work shows the simplicial stochastic SIR here defined differs significantly from the traditional Markovian SIR on networks, providing a richer structure to work with. Other possible lines of future work include: the further generalization of known results in the Markovian network SIR to the simplicial case; a deeper analysis of the simplicial stochastic SIR via theory of stochastic processes; exploration of other dynamics on networks for which a simplicial complex offers a suitable generalization; a study of how topological properties of the simplicial complex (e.g. homology) affects the epidemic process on it.

\section{Acknowledgments}
First author was fully supported by CONACyT scholarship number 779613.

\bibliographystyle{plainnat}
\bibliography{ref}      

\vspace{2cm}

\noindent Gerardo Palafox-Castillo \\
Facultad de Ingeniería Mecánica y Eléctrica \\
Universidad Autónoma de Nuevo León \\
San Nicolás de los Garza, Nuevo León, México \\
E-mail: \texttt{gerardo.palafoxcstl@uanl.edu.mx}\\

\vspace{1cm}
\noindent Arturo Berrones-Santos \\
Facultad de Ingeniería Mecánica y Eléctrica \\
Universidad Autónoma de Nuevo León \\
San Nicolás de los Garza, Nuevo León, México \\
E-mail:  \texttt{arturo.berronessn@uanl.edu.mx}\\
\end{document}